Robust Surface Hall Effect and Nonlocal Transport in SmB6: Indication for an Ideal Topological Insulator

D.J. Kim*, S. Thomas*, T. Grant, J. Botimer, Z. Fisk and Jing Xia

*Dept. of Physics and Astronomy, University of California, Irvine, California 92697, USA*
(* These authors contributed equally to this work.)

A topological insulator (TI) is an unusual quantum state in which the insulating bulk is topologically distinct from vacuum, resulting in a unique metallic surface that is robust against time-reversal invariant perturbations. When placed in a magnetic field, the two-dimensional surface necessarily leads to a surface Hall effect that is independent of sample thickness. Robust nonlocal transport through the highly conductive surface defies Ohm's law and could be useful for novel electronic devices. These surface transport properties, however, remains difficult to be isolated from the bulk in existing TI crystals ($Bi_2Se_3$, $Bi_2Te_3$ and $Sb_2Te_3$) due to impurity caused bulk conduction. We report in large crystals of topological Kondo insulator (TKI) candidate material $SmB_6$ the thickness-independent surface Hall effects and non-local transport, which are robust against perturbations including mechanical abrasion. These results serve as proof that at low temperatures SmB6 has a robust metallic surface that surrounds a truly insulating bulk, paving the way for transport studies of the surface state in this proposed TKI material.

Recently discovered [1-6] three-dimensional (3D) topological insulators (TI) have generated great excitement. Strong spin-orbit coupling in a TI gives rise to a non-trivial and robust conducting surface state [4,5], reminiscent to the edge channel [7,8] found in quantum Hall (QH) [9,10] and quantum spin Hall (QSH) [11,12] states. However, such surface transport properties has remained challenging to separate from residual bulk impurity conduction [13-16], promoting us to search for TI materials with a truly insulating bulk. More importantly, so far most theoretical and experimental efforts have been dedicated to materials with the underlying physics of non-interacting electrons [1,4,5,11]. A big question thus concerns the role of electron interaction and competing orders in new TI materials with strong correlation, from which new types of topological phases are expected to emerge [4,5].

Recent seminal theoretical works [17,18] have predicted in the strongly correlated material $SmB_6$ the existence of a topological insulator phase: topological Kondo insulator (TKI). $SmB_6$ is a heavy fermion material first studied 40 years ago [19]. In $SmB_6$ highly renormalized f-electrons, hybridized with conduction electrons, form a completely filled band of quasiparticles with a transport excitation gap $\Delta$ of about 40 Kelvin. However, its many exotic properties [20-22] still defy a satisfactory understanding within the framework of conventional insulators. One of these mysteries is the peculiar residual conduction at the lowest temperatures: behaving electronically like an insulator at high temperatures, at low temperature its resistance mysteriously saturates: a curiosity that is usually attributed to the existence of density of states within the band gap. According to the recent TKI theory [17,18,23,24], the hybridization and odd parity wavefunction lead to strong spin-orbit coupling in $SmB_6$ and gives rise to a topological surface state, which naturally explains the origin of the in-gap state and pinpoints its location to be on the surface of $SmB_6$. Our recent capacitance measurements on high quality $SmB_6$ crystals revealed intriguing anomalous capacitance effects [25] that could be explained by assuming the in-gap-states exist on the surface. In this paper we present evidences of robust surface Hall effect and nonlocal transport in high quality $SmB_6$ crystals of various geometries and from different growth batches. These results reveal in $SmB_6$ a perfectly insulating bulk and a robust conducting surface.

RESULTS

Hall effect measurements were carried out in wedge-shaped $SmB_6$ crystals. As depicted in the inset in Fig. 1(a), the sample is placed in a perpendicular magnetic field $\vec{B}$ and current $I$ flows between the two ends of the wedge. The Hall resistances $R_{xy} = V_{xy}/I$ are measured at different thicknesses $d$ to distinguish between surface and bulk conductions. For bulk conduction $R_{xy}/B \propto 1/d$, while $R_{xy}/B$ is $d$-independent if surface conduction dominates. The Hall voltage $V_{xy}$ is found to be linear with $B$ (Fig. 1(c) (d)) at small fields at all temperatures, but becomes significantly nonlinear around 5K, indicating a temperature regime of multichannel conduction. At high (20 K) or low (2K) temperatures, the extreme linearity of the Hall effect indicates single channel conduction, either it is the bulk or surface. For the simplest case of two independent channels with Hall coefficients $R_1, R_2$ and resistivity $\rho_1, \rho_1$, the Hall resistivity $\rho_{xy}$ at magnetic field $B$ is $\rho_{xy} = [(R_1\rho_1^2 + R_2\rho_2^2)B + R_1 R_2 (R_1 + R_2) B^3] / [(\rho_1 + \rho_2)^2 + (R_1 + R_2)^2 B^2]$

. Nonlinearity is expected at large $B$, but at small fields it simplifies to $\rho_{xy}/B = (R_1\rho_1^2 + R_2\rho_2^2)/(\rho_1 + \rho_2)^2$, which indeed gives thickness independent $R_{xy}/B$ if both channels are of surface nature. From $B < 1\,T$ data we extract the value $R_{xy}/B$ at various temperatures $T$. Representative results in sample *S1* are plotted in Fig. 1(a) for $d = 120, 270\ and\ 320\ \mu m$ respectively, showing clearly that while at high temperatures $R_{xy}/B$ differ at different $d$, they converge to a same universal value of $0.3\ \Omega/T$ below 4 Kelvin, in consistence with surface conduction. Since more than one surface channels may exist, as predicted by theory [23,24], it is difficulty to quantitatively extract the surface carrier density and mobility at this stage. Replotting the Hall resistance ratios $R_{xy}(d_1)/R_{xy}(d_2)$ in Fig. 1(b), we found these ratios to be equal to $d_2/d_1$ at high $T$ and become unity at low $T$, proving the crossover from 3D to 2D Hall effects when $T$ is lowered. The temperature dependence is well described by a two-channel (bulk and surface) conduction model in which the bulk carrier density decrease exponentially with temperature with an activation gap $\Delta= 38K$. Using this simple model, we could reproduce the curious "peak" in $R_{xy}/B$ at 4 $K$ (solid lines in Fig. 1(a)), which lacks [26] a good explanation until now.

Surface dominated conduction could also be demonstrated at zero magnetic field with so-called "nonlocal" transport, in the spirit of nonlocal transports experiments performed in QH [9,10] and QSH [11,12] states that have served as evidences [7,8] for the existence of the topological edge states that are one-dimensional analogues to the surface state in a TI. The highly metallic surface conduction in a TI would necessarily invalid Ohm's law and introduce large nonlocal voltages, which we have indeed found in $SmB_6$ samples. Fig. 2(a) shows a schematic of the nonlocal measurement in sample *S4*. Current $I_{16}$ flows between current leads 1 and 6 at the centers on opposite faces of the crystal. Contacts 2 and 3 are located close to contact 1 for the detection of "local" voltage $V_{23}$. Contacts 4 and 5 are put near the sample edge far away from the current leads to detect "nonlocal" voltage $V_{45}$. As shown in the inset in Fig. 2(b), in the case of bulk conduction, current will concentrate in the bulk near the current leads 1 and 6, resulting in negligibly small nonlocal voltage $V_{45} \ll V_{23}$. If surface conduction dominates, however, current will be forced to flow between contacts 1 and 6 via the surface, making $V_{45}$ large. Fig. 2(a) shows as a function of temperature the measured $V_{45}$ and $V_{23}$ divided by $I_{16}$, both agreeing qualitatively with our finite element simulations (Supplementary Information) incorporating the aforementioned simple model. The ratio $V_{45}/V_{23}$ is replotted in Fig. 2(b). At low temperature when surface conduction dominates the nonlocal voltage $V_{45}$ becomes large and even surpasses the local voltage $V_{23}$. Above $T = 5 K$ when bulk conduction dominates, the magnitude of $V_{45}/V_{23}$ is very small. The negative sign of $V_{45}/V_{23}$ is due to the small misalignment of contacts (Supplementary Information). The change of both magnitude and sign of $V_{45}/V_{23}$ is reproduced by our simulation and highlights the distinction between high $T$ bulk conduction and low $T$ surface conduction.

An important aspect of TI is the topological protection of the surface state against time-reversal invariant perturbations. The robustness against perturbation distinguishes a topological surface state from "accidental" surface conduction [4,5]. We found the above surface Hall effect and nonlocal transport are recurrent phenomena in various samples and are robust to chemical and mechanical sample treatments. As an illustration, we mechanically cut and scratched the surface of sample *S10* (Fig. 3 inset) and performed Hall measurements before and after surface abrasion. We found the low temperature Hall resistance $R_{xy}/B$ remain unchanged (Fig. 3(c)), indicating that the surface carrier density $n_S$ was not affected by abrasion. The abrasion does reduce the surface mobility $\mu_S$ though, as reflected by the height of $R_{xy}/B$ "peak" at $T = 4\ K$, presumably due to greatly enhanced geometric roughness. As shown in Fig. 3 (a) (b), the surface dominated conduction persists after abrasion as demonstrated by thickness-independent Hall effect below 4K.

DISCUSSION

The experimental identification of robust thickness-independent surface Hall effect and nonlocal transport serve as strong evidence that $SmB_6$ has a robust metallic surface state surrounding a truly insulating bulk. The characterization of energetics of the surface state and hence direct tests of the topological nature, however, awaits future investigations using energy and spin resolved techniques like ARPES [2,3] and STM [27,28]. Unlike weakly interacting TI materials, the strong electron correlation in $SmB_6$ could give rise to exotic emergent phases [4,5] with exciting new physics.

*Note added*: During initial submission of the manuscript, we became aware of a related work[3,29] in which evidence for surface conduction was provided in a $SmB_6$ sample with contacts arranged in a unique configuration, and a point contact measurement [9,10,30] that excludes the possibility that the in-gap state is located in the bulk of $SmB_6$.

METHODS

High quality $SmB_6$ crystals were grown using the aluminium flux method. The surfaces of these crystals were carefully etched using hydrochloric acid and then cleaned using solvents to remove possible oxide layer or aluminium residues. These crystals are then inspected using X-ray analysis to make sure $SmB_6$ is the only content. Samples used in the experiments were made from these crystals either by mechanical cleaving or polishing using polishing films containing diamond particles. The exposed surfaces are (100) planes. Gold and platinum wires are attached to the samples using micro spot welding and/or silver epoxy, with no discernable differences in measurements. Low frequency transport measurements were carried out in dilution fridges and helium cryostats using either standard low frequency (37 Hz) lock-in techniques with $50\ nA$ excitation currents or with a resistance bridge.

Acknowledgements We thank T.H. Geballe and A. Kapitulnik for useful discussions. This work was supported by UC Irvine CORCL Grant MIIG-2011-12-8, Sloan Research Fellowship (J. X.) and NSF grant #DMR-0801253.

Author Contributions J.X conceived the project. D.J.K, T.G. and Z.F. grew the crystals. D.J.K., S.T., J.B., and J.X. fabricated the samples and performed the measurements. T.G. performed X-ray analysis of the crystals. S.T. and J.X. performed the simulations. All authors analysed the data and wrote the manuscript.

Author Information Correspondence and requests for materials should be addressed to J.X. (xia.jing@uci.edu).

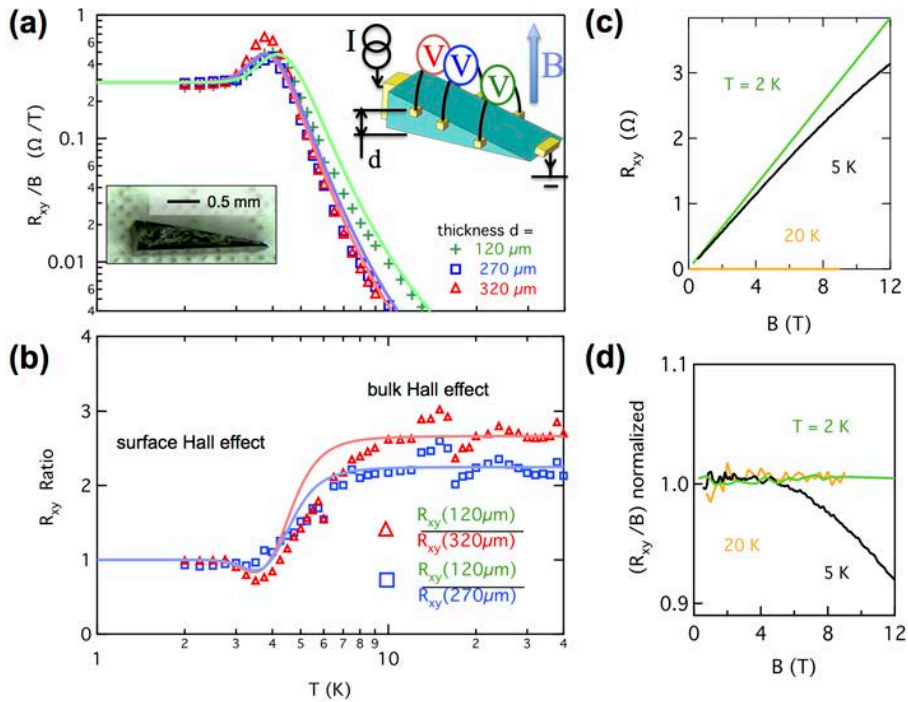

Figure 1 | Surface Hall effect. a, Markers, Hall resistances $R_{xy}$ divided by magnetic field $B$ versus temperature $T$ at three different thicknesses $d$ in a wedge shaped sample $S1$. Lines are simulations using a two conduction channel model (see text). Left inset, picture of the crystal before wiring. Right inset, measurement schematic. b, Markers, ratios between Hall resistances $R_{xy}$ at different $d$, showing the transition from bulk to surface conduction as temperature is lowered. Lines are calculated from simulations as in a. c, $R_{xy}$ versus $B$ at various $T$ for $d$ = 120 µm, showing nonlinearity at around 5 K. d, $R_{xy}/B$ normalized to small field values to demonstrate the nonlinearity.

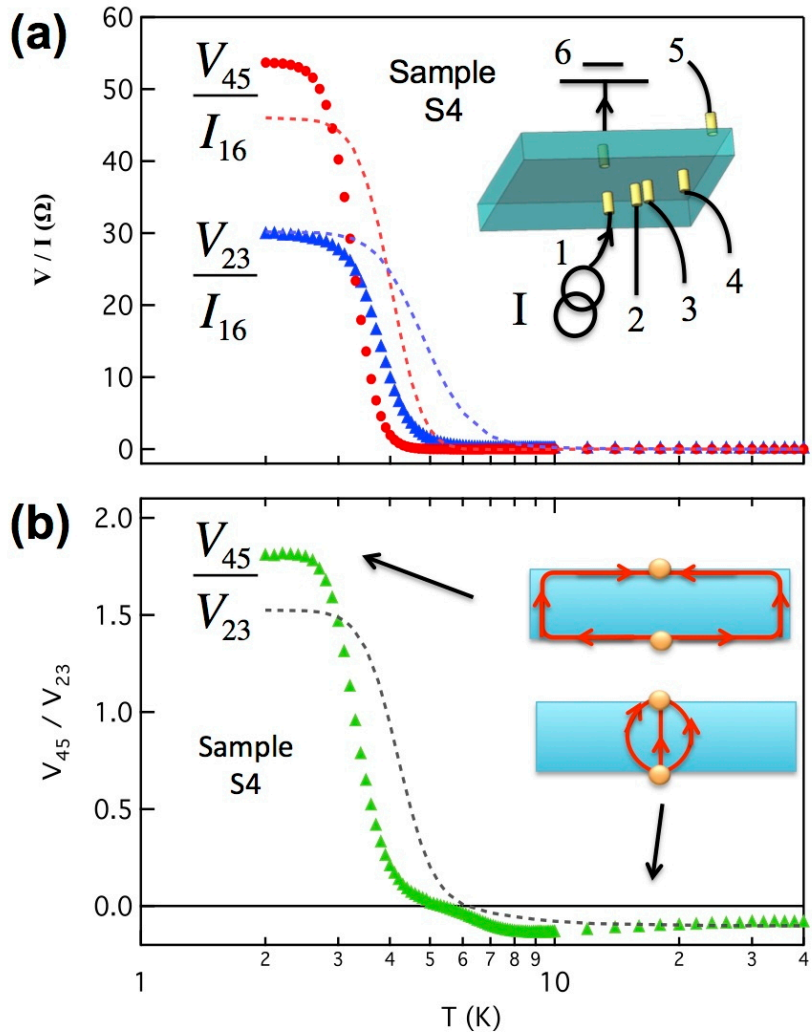

Figure 2 | Nonlocal transport due to surface conduction. a, Markers, nonlocal resistance $V_{45}/I_{16}$ and local resistance $V_{23}/I_{16}$ versus temperature $T$. Dashed lines are finite element simulations. Inset is a schematic of the measurement configuration on sample *S4*. b, Ratio between nonlocal and local voltages $V_{45}/V_{23}$ versus $T$. Dashed lines are finite element simulations. Inset, cartoons for current distribution in sample cross-section for surface and bulk dominated conductions.

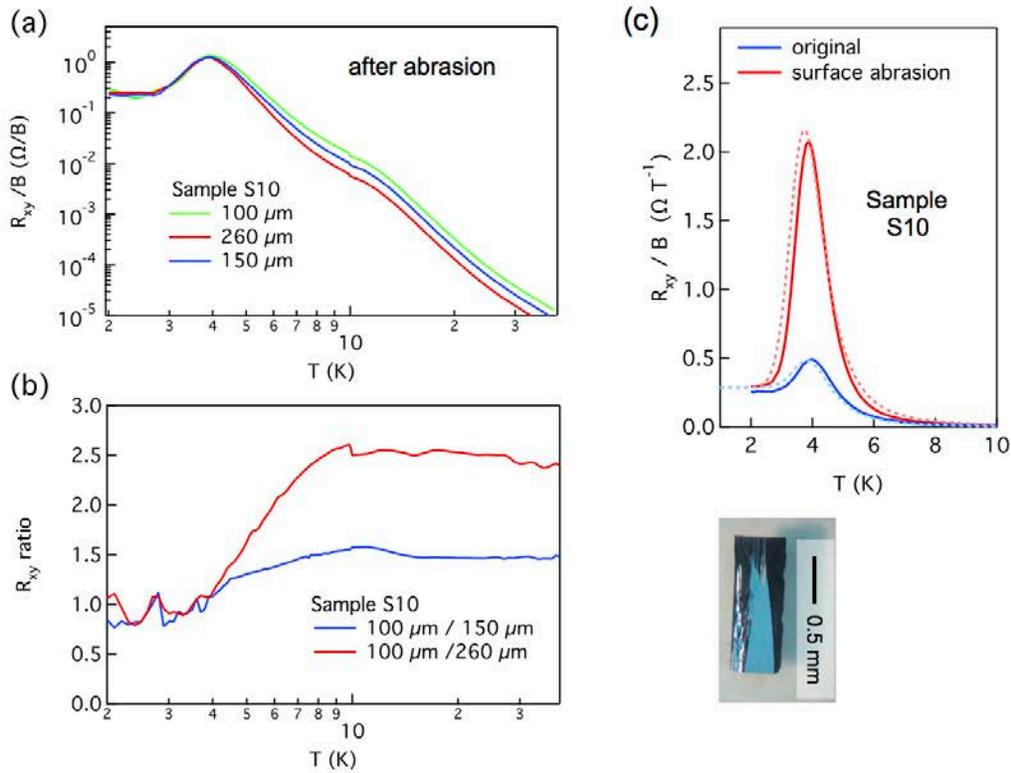

Figure 3 | Robustness of the surface state. a, Markers, Hall resistances $R_{xy}$ divided by magnetic field $B$ versus temperature $T$ at three different thicknesses $d$ in a wedge shaped sample *S10, after intentional chemical etching and cutting*. b, Markers, ratios between Hall resistances $R_{xy}$ at different $d$, showing the transition from bulk to surface conduction as temperature is lowered. c, Solid lines, Hall resistance $R_{xy}$ divided by magnetic field $B$ versus temperature $T$, *before and after surface abrasion*. Dashed lines are simulations assuming abrasion *only* reduces the effective surface mobility $\mu_S$. Inset, picture of sample S*10* during abrasion.

*Supplemental information:*

For the case of two independent channels with Hall coefficients $R_1, R_2$ and resistivity $\rho_1, \rho_1$, the Hall resistivity $\rho_{xy}$ at magnetic field $B$ is:

$$\rho_{xy} = [(R_1\rho_1^2 + R_2\rho_2^2)B + R_1 R_2 (R_1 + R_2) B^3] / [(\rho_1 + \rho_2)^2 + (R_1 + R_2)^2 B^2]$$

Nonlinearity is expected at large $B$, but at small fields it simplifies to:

$$\rho_{xy}/B = (R_1\rho_1^2 + R_2\rho_2^2)/(\rho_1 + \rho_2)^2.$$

To first order, we assume in $SmB_6$ the transport is governed by a temperature dependent bulk channel, and a temperature independent surface channel. Note that it is still possible to have multiple surface channels, thus we take the effective 2D resistivity as $\rho_S$, and Hall resistance as $R_S$, but can't infer carrier density or mobility until more information is known on the surface carrier types (electrons, holes or both). The bulk is a gapped insulator with an indirect activation gap $\Delta = 38\ K$, as calculated from temperature dependence of Hall effect at high temperatures. The bulk carrier density $n_B$ thus follows activation law of an insulator: $n_B = n_B^0 \exp(-\Delta/k_B T)$, where $k_B$ is the Boltzmann's constant and $n_B^0$ is a constant. In the simplified case of temperature independent mobility, this gives activated Hall coefficient $R_B = R_B^0 \exp(\Delta/k_B T)$ and resistivity $\rho_B = \rho_B^0 \exp(\Delta/k_B T)$.

For a sample with length, width and thickness of $L, w\ d$ respectively. The longitudinal resistance is just the parallel resistance of the surface and bulk channels:

$$R_{xx} = \frac{L}{w(2/\rho_S + d/\rho_B)} = \frac{L}{w(\frac{2}{\rho_S} + d/\rho_B^0 \exp(\Delta/k_B T))}$$

The factor 2 comes due to the fact there are bottom and top surfaces.

And the Hall resistance in the small field limit is:

$$R_{xy} = B\ \frac{2\frac{R_S}{\rho_S^2} + dR_B/\rho_B^2}{(2/\rho_S + d/\rho_B)^2} = B\ \frac{2\frac{R_S}{\rho_S^2} + dR_B^0 \exp(\Delta/k_B T)/(\rho_B^0 \exp(\Delta/k_B T))^2}{(\frac{2}{\rho_S} + d/\rho_B^0 \exp(\Delta/k_B T))^2} = B\ \frac{2\frac{R_S}{\rho_S^2} + \frac{dR_B^0}{(\rho_B^0)^2}\exp(-\Delta/k_B T)}{(\frac{2}{\rho_S} + \exp(\frac{\Delta}{k_B T})d/\rho_B^0)^2}$$

Using this simple model we could calculate temperature dependences of $R_{xx}$ and $R_{xy}/B$, e.g. as shown in Fig. S1. We could also perform finite element simulations for the transport in more complicated geometries. Fig. S2 and Fig. S3 show the simulated surface potential profiles at $300\ K$ and $0\ K$ for the nonlocal transport experiment in a sample with thin plate geometry. The simulation curves in Fig. 3 in the main text are generated using this finite element simulation.

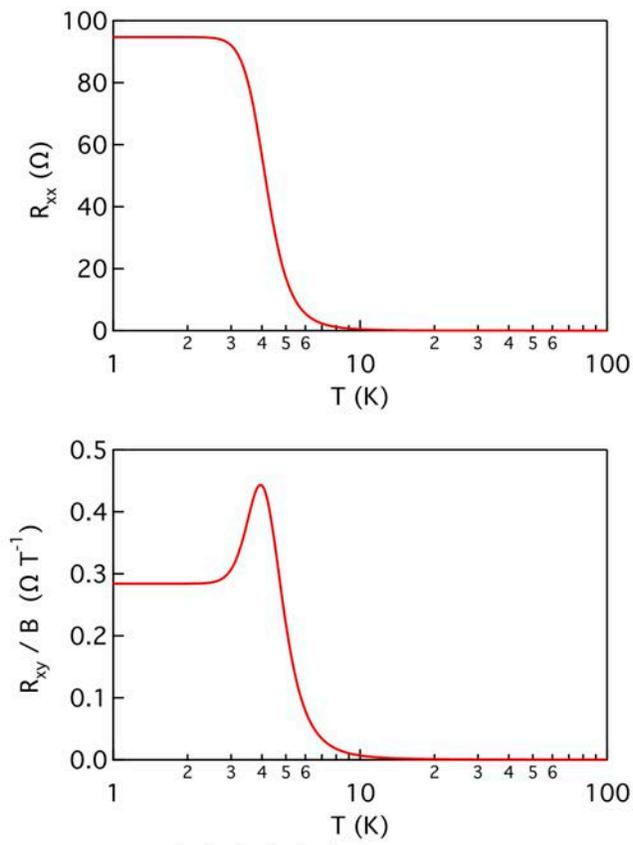

**Figure S1 | Simple Transport Model.  a, b,** Calculated $R_{xy}$ and $R_{xx}$ using the simple model of two channel conduction.

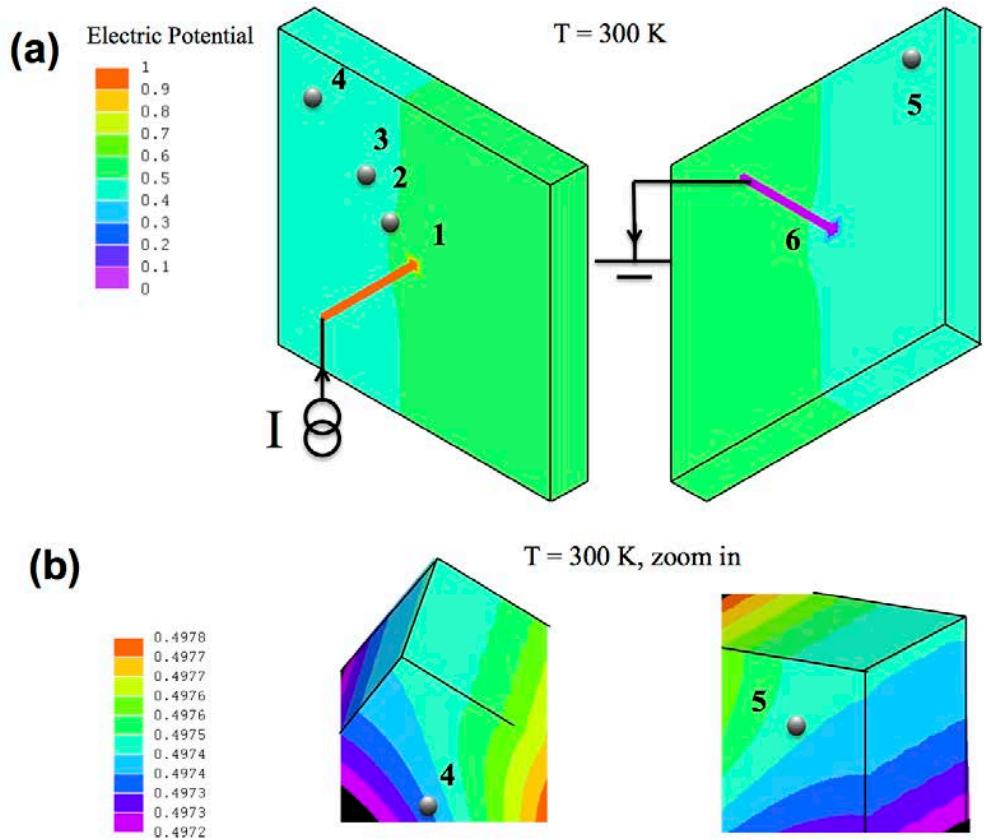

**Figure S2 | Simulation of Nonlocal Transport at 300 K. a,** Finite elements simulation of the surface potential profile in the nonlocal transport. As the conduction is dominated by the bulk, the nonlocal voltage $V_{45}$ is much smaller than $V_{23}$. **b,** Zoom in views for contact 4 and 5, showing $V_{45}$ is a small negative value due to slight misalignment of contacts.

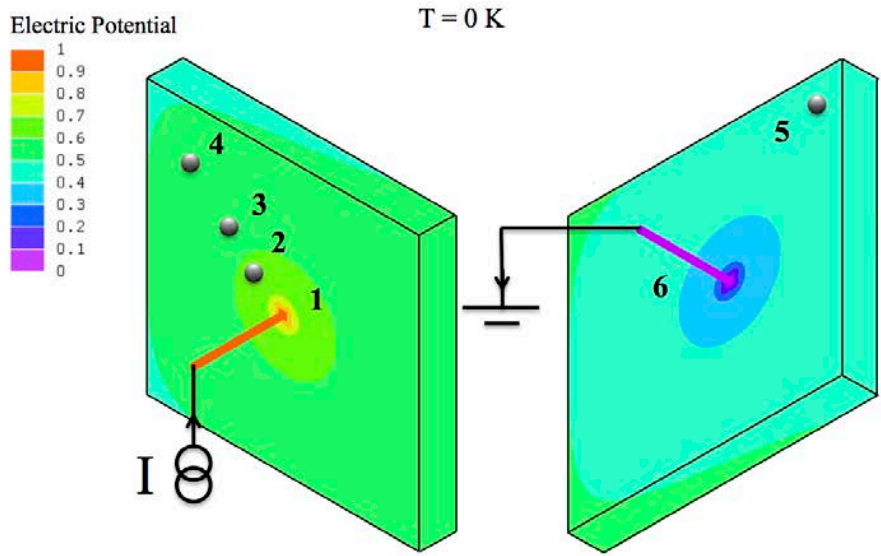

**Figure S3 | Simulation of Nonlocal Transport at 0 K. a,** Finite elements simulation of the surface potential profile in the nonlocal transport. Due to surface conduction, the nonlocal voltage $V_{45}$ is large and is comparable to $V_{23}$.